\documentclass[prl,aps,amsmath,twocolumn]{revtex4}
\usepackage{graphicx}
\usepackage{comment}
\usepackage{amsthm}
\usepackage{amsfonts}
\gdef\ffrac#1#2{\textstyle{\frac{#1}{#2}}\displaystyle}
\gdef\be{\begin{equation}}
\gdef\ee{\end{equation}}
\gdef\l{{l}}

\begin{document}
\title{Fluids in random media and dimensional augmentation}
\author{John Cardy}
\affiliation{All Souls College, University of Oxford OX1 4AL, UK\\
and Department of Physics, University of California, Berkeley CA 94720, USA.}
\email{john.cardy@all-souls.ox.ac.uk}
\begin{abstract}
We propose a solution to the puzzle of dimensional reduction in the random field Ising model, asking: to what random problem in $D=d+2$ dimensions does a pure system in $d$ dimensions correspond? For a continuum binary fluid and an Ising lattice gas, we prove that the mean density and other observables equal those of a similar model in $D$ dimensions, but with infinite range interactions and correlated disorder in the extra two dimensions. There is no conflict with rigorous results that the finite range model orders in $D=3$. Our arguments avoid the use of replicas and perturbative field theory, being based on convergent cluster expansions, which, for the lattice gas, may be extended to the critical point by the Lee-Yang theorem. Although our results may be derived using supersymmetry, they follow more directly from the matrix-tree theorem.
\end{abstract}

\maketitle
The subject of dimensional shift, in the context considered in this
paper, has a long and rather complex history.
That one physical theory in $D$ space dimensions should be related to another
in $d=D-2$ dimensions was first suggested in the context of the field-theoretic
formulation of the critical 
behavior of the Ising model in a quenched random magnetic field (RFIM) \cite{Aharony,Young}, where 
it was noticed that the most infrared singular Feynman diagrams in $6-\epsilon$ dimensions are equal to those of the non-random Wilson-Fisher fixed point in $4-\epsilon$ dimensions. This suggests that the lower critical dimension of the random field model, at and below which the system does not order, should be $D_l=3$. This dimensional reduction was later explained by Parisi and Sourlas \cite{ps1} in terms of an emergent supersymmetry  of the classical field equations. This adds two anti-commuting dimensions which, in integrals, cancel  with two of the $D$ commuting coordinates.

However it appears to contradict a heuristic argument due to Imry and Ma \cite{im} that $D_l=2$, later made rigorous by Imbrie \cite{imb}. Numerical studies \cite{num1,num2,num3,num4} suggest that dimensional reduction breaks down for $D<5$.
Over the years several explanations \cite{exp1,exp2,exp3,exp4,exp5,exp6} have been put forward to explain this discrepancy, one of which is that terms which break the supersymmetry become relevant \cite{susybr1,susybr2}.  More recently Kaviraj, Rychkov and Trevisani \cite{krt} confirmed this through an exhaustive analysis of a formulation \cite {jc2} of the replicated theory in which supersymmetry, and the terms which break it, are explicit. 

Meanwhile Parisi and Sourlas \cite{ps2} had applied their ideas to the problem of enumerating branched polymer (BP) configurations, and concluded
 that their critical properties in $D<8$ dimensions should the same as those of the Yang-Lee (YL) edge singularity of the Ising model in a purely imaginary field \cite{mef1}, or equivalently \cite{rep} the universal repulsive gas singularity at negative activity, in $D-2$ dimensions. This correspondence appears to hold down to $D=2$, in that series expansions for the branched polymer problem agree with exact results for the YL or repulsive gas problems. Kaviraj \em et al.\em~\cite{krt,kav2} found no evidence for relevant supersymmetry-breaking terms for this case.

Brydges and Imbrie \cite{bi,bi2,im2} devised a simple continuum model for branched polymers for which Parisi-Sourlas supersymmetry is exact and implies an equivalence to a repulsive gas model in two less dimensions. Their arguments hold to all orders in the activity expansion, which converges all the way up to the critical point. While this model is not generic since the interactions have to be fine-tuned, it is reassuring to have rigorous results which avoid the use of replicas and perturbative field theory. 

It is the purpose of the present work to apply these ideas to the RFIM, by inverting the question and asking instead:

\noindent\em To what, if any, quenched random system in $d+2$ dimensions does a nonrandom finite range model in $d$ dimensions correspond?\em

\noindent In order to make rigorous statements we eschew the replica trick or perturbative field theory, instead using cluster expansions on fluid models of Ising criticality, which yield convergent power series in the activity $z$. The first is a purely repulsive binary fluid which, in the absence of randomness, exhibits phase separation with a critical endpoint in the Ising universality class. That such a fluid in a medium with random affinity for one component over the other should correspond to the RFIM was first suggested by de Gennes \cite{dg}, with some later numerical and experimental support \cite{vink}. The advantage of this model is that it has a simple Mayer cluster expansion. However its convergence is controlled by the repulsive gas singularity at negative $z$, and so cannot straightforwardly be used to draw conclusions about the physical critical point which is further from the origin. The second is the well-known lattice gas representation of the Ising model, in which the Lee-Yang theorem \cite{yl1,yl2} allows the expansion in $z$ to be continued up to the Yang-Lee circle, where the density of zeroes is known to determine the critical behavior for $z>0$. 

\begin{figure}
\includegraphics[width=0.4\textwidth]{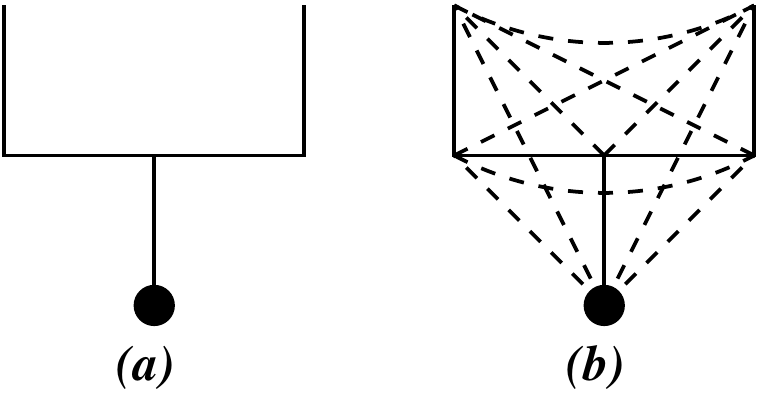}
\caption{\label{fig1}\em(a) \em a rooted tree; \em(b) \em a decorated rooted tree. In $D$ dimensions, the solid lines represent the Mayer functions and the dashed lines the disorder average.}
\end{figure}

In both the field theoretic approach \cite{ps1} and Brydges and Imbrie \cite{bi,bi2,im2}, integration over the anticommuting coordinates gives a sum over \em tree \em diagrams (see Fig.~\ref{fig1}) with certain weights. 
For Parisi and Sourlas' discussion of the RFIM \cite{ps1} these are the perturbative solution of the classical field equations. For their analysis of BP \cite{ps2} the Feynman diagrams themselves approximate the polymer configurations \cite{shap}.
For Brydges and Imbrie \cite{bi}  the trees give the partition sum of a microscopic model for BPs.
In this work, by contrast, we interpret the trees as a \em subset \em of the diagrams in the cluster expansion of an interacting fluid in $D$ dimensions. 
In fact the integration gives \em decorated \em trees, with multiplicative weights for every pair of vertices in the tree, whether adjacent on the tree or not. We interpret these  as the result of performing the quenched average over a random potential. 
\vspace{-2em}
\subsection*{Description of the models and cluster expansion.}
\vspace{-1em}
\noindent\em Binary fluid. \em The species are labeled by $a,b=1,2$ with activities $z_a$ and 2-body repulsive potentials $V_{ab}(r^2)$ with $V_{ab}\geq0$ and $V'_{ab}\leq0$. For convenience we take these all to have the same $r$-dependence up to multiplicative factors.
Phase separation at zero temperature then occurs if $2V_{12}>V_{11}+V_{22}$. The grand partition function is
\be\label{gp}
\sum_{N_a=0}^\infty\prod_a\frac{z_a^{N_a}}{N_a!}\prod_{i_a=1}^{N_a}\int e^{-\frac12\sum_{bc;j_bk_c}V_{bc}((r_{j_b}-r_{k_c})^2)}d^dr_{i_a}\,,
\ee
where $1\leq i_a\leq N_a$ labels the particles of species $a$, and $r_{a_i}$ are their positions, and so on.
In order to study the critical point it is useful, but not necessary, to consider the symmetric case $V_{11}=V_{22}$, $z_1=z_2$, which explicitly exhibits the Ising ${\mathbb Z}_2$ symmetry \cite{WR}. $V_{ab}$ then decomposes into even/odd subspaces, with eigenvalues $V_+>0$ and $V_-<0$.
In the absence of random fields, in mean field theory
there are two critical points: (i) one with $z<0$, where $\phi_+$ is critical. This is the \em repulsive gas singularity \em\cite{rep}. Eliminating the non-critical $\phi_-$ modes leads to a cubic field theory with imaginary coupling for the $\phi_+$ fluctuations, showing it to be in the same universality class as the Yang-Lee edge singularity \cite{mef1}; (ii) with $z>0$, where $\phi_-$ is critical. Eliminating the massive $\phi_+$ field then gives the usual Landau-Ginzburg-Wilson $\phi_-^4$ free energy. The random field $h_-=h_1-h_2$ then couples to the order parameter $\phi_-$, while $h_+$ couples to the massive mode.
The perturbative analysis of this model using replicas \cite{Aharony,Young,ps1} then leads to the dimensional reduction hypothesis and the aforementioned puzzle of the lower critical dimension. However, we shall not follow this path, instead using cluster expansion methods directly on (\ref{gp}). These proceed by writing $V_{ab}(r^2)=1+f_{ab}(r^2)$ and expanding (\ref{gp}) in powers of $f_{ab}$. Each term corresponds to a spanning subgraph of the complete graph with $N_1+N_2$ vertices, labeled by $r\in{\mathbb R}^d$ and colored $a=1$ or $2$, each edge of the subgraph being allocated a factor $f_{ab}$. The mean density $\langle n_a(0)\rangle$ is given by the sum over \em connected \em spanning subgraphs, or clusters $C$, with an $a$ vertex \em rooted \em at $r=0$. Each cluster integral has the form (suppressing temporarily the species labels) $C_d=\int' \prod_{jk\in e(C)}f(r_{jk}^2)\prod_id^dr_i$ where $e(C)$ is the edge set of $C$, and the prime on the integration indicates that $r_1$ is to be set to $0$.

\noindent\em Lattice gas. \em  The partition function is
\be\label{lg}
\sum_{n(r)=0,1}\prod_rz(r)^{n(r)}e^{-\frac12\sum_{rr'}n(r)V((r-r')^2)n(r')}\,,
\ee
 where $r,r'\in{\mathbb Z}^d$ and $V\leq0$, which maps exactly onto a ferromagnetic Ising model with spins $s(r)=2n(r)-1=\pm1$. On a finite graph (\ref{lg}) is a multinomial of degree one in each of the $z(r)$, and when these are all equal is a polynomial satisfying the Lee-Yang theorem \cite{yl1,yl2} that its zeroes all lie on a circle.  (\ref{lg}) may be transformed into a (lattice) field theory \cite{ftex}
\be\label{ft2}
\int[d\phi]e^{\frac12\sum_{rr'}\phi(r)V^{-1}((r-r')^2)\phi(r)+\sum_r\log(1+z(r)e^{\phi(r)})}\,.
\ee
Expanding the logarithm then the exponential, and using Wick's theorem gives
\begin{eqnarray}
\sum_{N=0}^\infty\frac1{N!}\sum_{r_i\in{\mathbb Z}^d}\sum_{p_i=1}^\infty\Big(\prod_{i=1}^N\frac{(-1)^{p_i-1}}{p_i}z(r_i)^{p_i}
e^{-\frac12p_i^2V(0)}\Big)\nonumber\\
\times\Big(\prod_{1\leq j<k\leq N}e^{-p_jp_kV(r_{jk}^2)}\Big)\,,\label{xii}
\end{eqnarray}
in which each vertex is now labeled by $r_i\in{\mathbb Z}^d$ and $p_i\in{\mathbb N}^+$. (\ref{xii}) has the property that the sums over the $(r_i,p_i)$ are unrestricted and therefore, as for the binary repulsive gas, a simple Mayer expansion in $f_{p_jp_k}(r_{jk}^2)=e^{-p_jp_kV(r_{jk}^2)}-1$ holds for the local density as a sum of connected rooted diagrams.  Note that to any fixed order in $z$, the sums over $N$ and the $p_i$ may be truncated.
\vspace{-2em}
\subsection*{Statement of the main result.}
\vspace{-1em}
We denote the augmented coordinates by $\rho_\mu\in{\mathbb R}^2$. A prime denotes the derivative with respect to $\rho^2$.\\
\noindent{\bf Theorem (Dimensional Augmentation.)} 
\em The mean densities of a non-random binary fluid, or of a lattice gas, in ${\mathbb R}^d$ (resp.~${\mathbb Z}^d$) with 2-body  potentials $V(r,r';0)$ are equal, to all orders in the activity $z$, to those for similar models in ${\mathbb R}^d\otimes{\mathbb R}^2$, with potentials
$-(1/\pi\l^2)V'(r,r';\rho^2/\l^2)$, when averaged over a random field drawn from a gaussian distribution with covariance $-V(r,r';\rho^2/\l^2)$, in the limit as $\l\to\infty$.\em

In fact ${\mathbb R}^d$ may be replaced by an arbitrary measurable continuous or discrete set $\cal G$, with arbitrary fixed potentials inserted: only the rotational and translational symmetries in the extra two dimensions are important. This allows the extension to arbitrary correlation functions and other geometries. 
\vspace{-2em}
\subsection*{Outline of the proof.}
\vspace{-1em}
We first consider the binary fluid, 
using the following identity: 
\be\label{lemma}
C_d=\!\!\sum_{T\subseteq C}\int'\!\!\! \prod_{jk\notin e(T)}\!\!f(r_{jk}^2)\!\!\!\!\prod_{lm\in e(T)}\!\!(-\ffrac1\pi f'(r_{lm}^2))\prod_id^{D}r_i\,,
\ee
where the sum is over all connected spanning rooted tree subgraphs $T$ of $C$, and $\int'$ indicates that $r_1$ is not integrated over, but set to $0$. This follows by writing $f(r_{jk}^2)=\int\tilde f(\alpha_{jk})e^{-\alpha_{jk}r_{jk}^2}d\alpha_{jk}$ so that, for fixed $\{\alpha_{jk}\}$, the integrals over the $\{r_i\}$ have the form $\int' e^{-\sum_{lm}{\mathbf r}_lA_{lm}{\mathbf r}_m}d^dr_i$
$\propto(\det A')^{-d/2}=(\det A')(\det A')^{-D/2}$, where $A_{lm}=A_{ml}=\alpha_{lm}$ for $l<m$ and $A_{mm}=-\sum_{l<m}\alpha_{lm}$, and $A'$ is formed by deleting the first row and column of $A$. This is of precisely the form to apply the matrix-tree theorem \cite{kirch}, which asserts that $\det A'$ is given by the sum over spanning trees weighted by the product of the $\alpha_{lm}$ on each edge of the tree. These factors are then equivalent to replacing $f\to f'$ on these edges, giving (\ref{lemma}).

The next step is to sum over all clusters $C$ containing a fixed tree $T$. Since each edge $\notin e(T)$ is either $\in C$ or $\notin C$, this is equivalent to replacing $f(r_{jk}^2)\to1+f(r_{jk}^2)=e^{- V(r_{jk}^2)}$ on each such edge.  Noting that $f'=-V'e^{-V}$, and restoring the species labels, we therefore have
\be\label{treesum}
\langle n_a\rangle_d=\sum_T\prod_{bc}z_b^{N_b}\!\!\!\int'\!\!\!\!\!\prod_{jk\in e(T)}\!\!\!\!\ffrac1\pi V'_{bc}(r_{jk}^2)\prod_{\forall jk}e^{-V_{bc}(r_{jk}^2)}\prod_i d^Dr_i\,,
\ee
which is a multispecies version of the main result of Brydges and Imbrie \cite{bi}. 
It may also be derived using supersymmetry, adding two further anticommuting coordinates and showing that (a) their Berezin integrals in cluster diagrams cancel those over the $\rho_\mu$; (b) the non-zero diagrams are (decorated) trees. The above authors interpret (\ref{treesum}) as the partition function for branched polymers, with an attractive weight with short-range repulsion $-V'e^{-V}$ between neighbors on the tree, and a repulsive weight $e^{-V}$ between all other pairs. The extra minus signs are absorbed into the activities $z_a$, so that BPs in $D$ dimensions with $z_a>0$ are mapped into the repulsive gas at negative activity. The multispecies version adds nothing new, since it is the combination $\phi_+$ which becomes critical there.

Alternatively we may consider (\ref{treesum}) as arising from the expression 
\be\label{ran}
\sum_C\prod_az_a^{N_a}\!\!\!\int'\!\!\!\prod_{jk\in e_{bc}(C)}\!\!\widetilde f_{bc}(r_{jk}^2+\rho_{jk}^2))\prod_i e^{h_a(r_i,\rho_i)}d^dr_id^2\rho_i
\ee
for the density of a fluid in a random medium in $D$ dimensions, averaged over a gaussian distribution for $h_a$ with covariance as stated, and Mayer functions $\widetilde f_{bc}=(1/\pi\l^2)V'_{bc}(r^2+\rho^2/\l^2)$, corresponding again to a repulsive gas exhibiting phase separation. Note that we do not need replicas because we are averaging the density directly. The limit $\l\to\infty$ restricts the sum over clusters $C$ to the subset of trees, since, on integration over the $\rho_i$, the latter are independent of $\l$, while each loop acquires a factor $\l^{-2}$ \cite{trees,d}. The relative error in neglecting the loop diagrams can be shown to be $\sim c_N\l^{-2}$ where $c_N$ grows no faster than an exponential.
This then establishes the theorem, as long as the pure cluster expansion converges. This is known to be true for sufficiently small $z_a$ \cite{conv,blh}, and should hold all the way up to the repulsive singularity, which is the closest to the origin. It also implies that any numerical extraction of the critical behavior at $z>0$ from the power series using, for example, Pad\'e approximants, will lead to identical results for the two systems.
Note that while the covariance $\overline{h_-h_-}=-V_-$ of the random field coupling to the order parameter $n_-$ is positive, that of $h_+$ is negative. This may instead be regarded as a purely imaginary field of positive covariance coupling to $n_+$, which makes sense because the order parameter at the repulsive gas singularity is the imaginary part of the fluctuation in $n_+$.
  
For the lattice gas, the argument is similar, with $\widetilde f_{p_jp_k}=e^{-p_jp_k\widetilde V}-1\sim\l^{-2}p_jp_kV'(r^2+\rho^2/\l^2)$. An important check is that, because of the $\l\to\infty$ limit,  $\widetilde V
\to0$ and so is independent of the $p_i$. The cluster expansion converges for sufficiently small $|z|$. However since it must reproduce the exact expansion in $z$ to any finite order, it follows from the Lee-Yang theorem \cite{yl2} applied to the pure $d$-dimensional model that it converges everywhere within the Yang-Lee circle. Thus the quenched average mean magnetization of the (long-range) random field $D$-dimensional model must also have this property \cite{griff}. In particular, they must have the same density of zeroes on the circle, and the same critical behavior at the Yang-Lee edge. Since this controls the physical critical behavior on the real $z$ axis as the critical temperature is approached, the pure $d$-dimensional Ising model and the $d+2$-dimensional random field Ising model with infinite-range interactions and correlated impurities have identical critical behaviors. 
\vspace{-2em}
\subsection*{Discussion.}
\vspace{-1em}
We have given a concrete answer to the question posed at the beginning: a non-random fluid in $d$ dimensions with finite range 2-body potentials is precisely equivalent to one in a quenched random medium in $d+2$ dimensions, in the limit when the range $\l$ of the interactions and correlated randomness in the two extra dimensions tends to infinity. It is \em not \em precisely equivalent to a fluid with only finite range interactions and impurity correlations, to which the arguments \cite{im,imb} that $D_l=2$ apply. 

However the present study does not immediately shed light on why dimensional reduction fails for $D<5$ in short range models. 
SUSY implies dimensional reduction even non-perturbatively \cite{jcsusy}, so presumably interactions which break it explicitly should become relevant at the SUSY fixed point, as concluded in Ref.~\cite{krt}. In the random model on which we land through dimensional augmentation, SUSY is explicitly realized in two ways: through tree dominance of the cluster expansion, which we have argued may be realized by taking the $\l\to\infty$ limit, but also through an explicit fine-tuning between the random field covariance $\propto-V$ and the interaction potential $\propto-V'$.

The relation of the first with the earlier field theory approaches may be seen in the $D$-dimensional version of (\ref{ft2}) by rescaling $\rho\to\l\rho$ and $V\to\l^{2}$, which modifies the measure to $e^{-\l^2{\cal H}}$. The limit $\l\to\infty$ then formally leads to the classical equations, the tree Feynman diagrams \cite{trees} and conjectured Parisi-Sourlas supersymmetry and dimensional reduction. In fact $\l^{-2}$ may be identified as the dangerously irrelevant variable in that approach, which is responsible for the breaking of hyperscaling. It is difficult to see how such a mechanism could fail as $D$ is lowered.  More likely is that the breaking of fine-tuning becomes relevant. It would be interesting to carry out numerical simulations on finite-range models in lower dimensions with fine-tuning of the random field covariance. 

Finally, we have pointed out that dimensional shift by two, which may be seen as a consequence of supersymmetry \cite{ps1,ps2}, also follows from Kirchhoff's theorem \cite{kirch} which preceded this by some 132 years \cite{discussion}.
 
\noindent\em Acknowledgements. \em 
The author thanks J.~Imbrie, S.~Rychkov and K.~Wiese for comments on earlier versions of this paper. This work was carried out while the author was affiliated with the Department of Physics, University of California, Berkeley and he thanks, in particular, J.~Moore for arranging this.


\begin{thebibliography}{99}
\bibitem{ftex} Note that we use this only as a means of generating a cluster expansion.
\bibitem{d} If ${\cal G}={\mathbb R}^d$ we could rescale all $D$ dimensions isotropically. However this loses the generality of the result, as well as the obvious extension to $\rho$-independent correlation functions.
\end{thebibliography}
\end{document}